\documentclass[10pt,journal,letterpaper]{IEEEtran}

\usepackage{amsmath,cite,psfrag,subfigure,graphicx,color,amssymb}
\usepackage{algorithm}
\usepackage{algorithmic}
\begin{document}
\graphicspath{{figures/}}
\newcommand{\sps}{\scriptsize}
\newcommand{\eqs}{\footnotesize}
\title{Statistical Speech Model Description
 with  VMF Mixture Model}

\author{Zhanyu~Ma and~Arne~Leijon
\IEEEcompsocitemizethanks{\IEEEcompsocthanksitem Zhanyu Ma is with Pattern Recognition and Intelligent System Lab., Beijing University of Posts and Telecommunications, Beijing 100876, China.
\IEEEcompsocthanksitem Arne Leijon is with School of Electrical Engineering, KTH - Royal Institute of
Technology, SE-100 44 Stockholm, Sweden.
}}

\IEEEcompsoctitleabstractindextext{
\begin{abstract}
Efficient quantization of the linear predictive coding (LPC) parameters plays a key role in parametric speech coding. The line spectral frequency (LSF) representation of the LPC parameters has found its applications in speech model quantization. In practical implementation of vector quantization (VQ), probability density function (PDF)-optimized VQ has been shown to be more efficient than the VQ based on training data. 
In this paper, we present the LSF parameters by a unit vector form, which has directional characteristics. The underlying distribution of this unit vector variable is modeled by a von Mises-Fisher mixture model (VMM). With the high rate theory, the optimal inter-component bit allocation strategy is proposed and the distortion-rate (D-R) relation is derived for the VMM based-VQ (VVQ). Experimental results show that the VVQ outperforms our recently introduced DVQ and the conventional GVQ.
\end{abstract}

}

\maketitle

\IEEEdisplaynotcompsoctitleabstractindextext

\IEEEpeerreviewmaketitle

\section{Introduction}
\IEEEPARstart{Q}{uantization} of the line predictive coding (LPC) model is ubiquitously applied in speech coding~\cite{Paliwal1995,Kleijn2003,Ma2012,Ma2014}. The line spectral frequency (LSF)~\cite{Itakura1975} presentation of the LPC model is the commonly used one in quantization~\cite{Paliwal1993,Paliwal1995} because of its relatively uniform spectral sensitivity~\cite{Li1999}. Efficient quantization methods for the LSF parameters have been studied intensively in the literature (see~\emph{e.g.},~\cite{Paliwal1993,Hedelin2000,Subramaniam2003,Ma2013}). Among these methods, the probability density function (PDF)-optimized vector quantization (VQ) scheme has been shown to be superior to those based on training data~\cite{Hedelin2000,Subramaniam2003}. In PDF-optimized VQ, the underlying distribution of the LSF parameters is described by a statistical parametric model,~\emph{e.g.}, Gaussian mixture model (GMM)~\cite{Hedelin2000}. Once this model is obtained, the codebook can be either trained by using a sufficient amount of data (theoretically infinitely large) generated from the obtained model or calculated theoretically. Thus PDF-optimized VQ can prevent the codebook from overfitting to the training data, and hence the performance of VQ can be significantly improved~\cite{Hedelin2000,Subramaniam2003}.

Statistical modeling plays an important role in PDF-optimized VQ, hence in the literature, several studies have been conducted to seek an effective model to explicitly capture the statistical properties of the LSF parameters or its corresponding transformations. A frequently used method is the GMM-based VQ (GVQ), which models the LSF parameters' distribution with a GMM~\cite{Hedelin2000,Subramaniam2003}. By recognizing the bounded property (all the LSF parameters are placed in the interval $(0,\pi)$), Lindblom and Samuelsson~\cite{Lindblom2003} proposed a bounded GVQ scheme by truncating and renormalizing the standard Gaussian distribution. In previous work, the LSF parameters were linearly scaled into the interval $(0,1)$. Authors introduced a beta mixture model (BMM)-based VQ scheme, which took into account the bounded support nature of the LSF parameters. As the LSF parameters are also strictly ordered, a Dirichlet mixture model (DMM)-based VQ (DVQ) scheme was recently presented to explicitly utilize both the bounded and the ordering properties~\cite{Ma2013}. In the DVQ scheme, the LSF parameters were transformed linearly to the $\Delta$LSF parameters. Modeling the underlying distribution of the $\Delta$LSF parameters with a DMM yields better distortion-rate (D-R) relation than those obtained by modeling the LSF parameters with a GMM~\cite{Subramaniam2003,Ma2014,Chatterjee2008} and a BMM. Hence, the practical quantization performance was also improved significantly~\cite{Ma2013}. Previous studies suggest the fact that transforming the LSF parameters into some other form and applying a suitable statistical model to efficiently describe the distribution can potentially benefit the practical quantization~\cite{Ma2013}.

In this letter, we study the high rate D-R performance of the LSF parameter by using the recently proposed square-root $\Delta$LSF (SR$\Delta$LSF) representation. This representation is obtained by taking the positive square-root of the $\Delta$LSF parameters. By concatenating a redundant element to the end of the SR$\Delta$LSF parameter, a unit vector that contains only positive elements is obtained. Geometrically, this unit vector has directional characteristics and is distributed on the hypersphere with center at the origin. For such unit vector, the von Mises-Fisher (vMF) distribution is an ideal and widely used statistical model to describe the underlying distribution~\cite{Mardia2000}. One application domain of vMF distribution is in information retrieval where the cosine similarity is an effective measure of similarity for analyzing text documents~\cite{Banerjee2005}. Another application domain of this distribution is in bioinformatics (e.g.,~\cite{Banerjee2005}) and collaborative filtering (e.g.,~\cite{Sarwar01}) in which the Pearson correlation coefficient serves as the similarity measure. More recently, Taghia et al. proposed a text-independent speaker identification system based on modeling the underlying distribution of SR$\Delta$LSF parameters by a mixture of vMF distributions. Here, we model the underlying distribution of the SR$\Delta$LSF parameters by a VMM and propose a VMM-based VQ (VVQ) scheme. According to the high rate quantization theory~\cite{Kleijn2010}, the D-R relation can be analytically derived for a single vMF distribution with constrained entropy. Based on the high rate theory, the optimal inter-component bit allocation strategy is proposed. Finally, the D-R performance for the overall VVQ is derived. Compared with the recently presented DVQ and the conventionally used GVQ, the VVQ shows convincing improvement. Hence, it potentially permits better practical quantization performance.

The remaining parts are organized as follows. In section~\ref{sec: Parameter Representations}, different representations of the LSF parameters are introduced. We briefly review the vMF distribution and the corresponding parameter estimation methods in section~\ref{sec: VMM}. A PDF-optimized VQ based on VMM is proposed in section~\ref{sec: PDF-optimized VQ} and the experimental results are shown in section~\ref{sec: Experiments}. Finally, we draw some conclusions and discuss future work in section~\ref{sec: Conclusion} and~\ref{sec: Future Work}.

\section{LSF, $\Delta$LSF, and SR$\Delta$LSF}
\label{sec: Parameter Representations}
\subsection{Representations}
The LSF parameters are widely used in speech coding due to the advantage over some other forms of representations (such as LARs,
ASRCs). The LSF parameters with dimensionality $K$ are defined as
\begin{equation}
\eqs
\mathbf{s} = [s_1,s_2,\ldots,s_K]^{\text{T}},
\end{equation}
which are interleaved on the unit circle~\cite{Itakura1975}.

By recognizing that the LSF parameters are in the interval $(0,\pi)$ and are
strictly ordered, we proposed a particular representation of LSF parameters called $\Delta$LSF for the purpose of LSF quantization~\cite{Ma2013}.
The $\Delta$LSF $\mathbf{v}$ parameters in represented as
\begin{equation}
\eqs
\label{TransformationFuncA}
\mathbf{v} = \varphi(\mathbf{s})=[s_1,s_2-s_1,\ldots,s_K-s_{K-1}]^{\text{T}}.
\end{equation}

Another representation of the LSF parameters were introduced, which took the square-root of the $\Delta$LSF parameters. Hence, a $K$-dimensional SR$\Delta$LSF $\mathbf{x}$ parameters can be obtained as
\begin{equation}
\eqs
\mathbf{x}=\phi(\mathbf{v})=\mathbf{v}^{\frac{1}{2}}=[\sqrt{v_1},\sqrt{v_2},\ldots,\sqrt{v_K}]^{\text{T}}.
\end{equation}
In, we modeled the underlying distribution of the SR$\Delta$LSF by a $(K+1)$-variate VMM and proposed a text-independent speaker identification system based on the SR$\Delta$LSF representation, which showed competitive performance compared to the benchmark approach.
\subsection{Distortion Transformation}
When getting the SR$\Delta$LSF parameters from the $\Delta$LSF parameters, the parameter space is wrapped. Hence, we study the distortion transformation between the SR$\Delta$LSF and the $\Delta$LSF spaces in this section.


Denote the PDFs of $\mathbf{v}$ and $\mathbf{x}$ as $g(\mathbf{v})$ and $f(\mathbf{x})$, respectively. Assuming that the $K$-dimensional SR$\Delta$LSF space is divided into $J$ cells and with the optimal lattice quantizer, the overall quantization distortion (using the square error as the criterion) for $\mathbf{x}$ can be
written as~\cite{Kleijn2010}
\begin{equation}
\eqs
D_\mathbf{x} = \sum_{j=1}^J \int_{\mathcal{V}_\mathbf{j,x}} \|\mathbf{x} - \mathbf{x}_j\|^2 f(\mathbf{x})d\mathbf{x}\approx \frac{1}{V_{\mathbf{x}}}\int_{\mathcal{V}_\mathbf{x}}\mathbf{e}^{\text{T}}\mathbf{e}d\mathbf{e}
\end{equation}
where
$\mathbf{\mathbf{e}}=\mathbf{\mathbf{x}}-\mathbf{\mathbf{\widehat{x}}}$
denotes the quantization error and all the cells
$\mathcal{V}_\mathbf{j,x}$ are of identical shape according to
Gersho conjecture~\cite{Kleijn2010}. The mapping $\phi$
from $\Delta$LSF space to SR$\Delta$LSF space changes the distortion per
cell in the $\Delta$LSF domain at $\mathbf{v}$ as
$\frac{1}{V_{\mathbf{v}}}\int_{\mathcal{V}_\mathbf{v}}\mathbf{e}^T\mathcal{J}_{\phi}(\mathbf{{\mathbf{v}}})^T\mathcal{J}_{\phi}(\mathbf{{\mathbf{v}}})\mathbf{e}d\mathbf{e}$,
where $\mathcal{J}_{\phi}(\mathbf{{\mathbf{v}}})$ is the
Jacobian matrix
\begin{equation}
\eqs
\mathcal{J}_{\phi}(\mathbf{{\mathbf{v}}})_{i,j}=\left\{ \begin{array}{cc}
\frac{\partial {\phi^{-1}}(\mathbf{x})_i}{\partial x_j}\mid_{\mathbf{x}=\phi(\mathbf{v})}=2 \sqrt{v_i} & i=j\\
0 & i\neq j
\end{array}\right ..
\end{equation}
Then the overall
quantization distortion transformation between $\mathbf{x}$ and
$\mathbf{v}$ can be denoted as
\begin{equation}
\label{DistortionTransformation}
\eqs
\begin{split}
D_\mathbf{v} =&\int_{\mathcal{V}_\mathbf{v}}g(\mathbf{v})\frac{1}{V_{\mathbf{x}}}\int_{\mathcal{V}_\mathbf{x}}\mathbf{e}^\mathrm{T}\mathcal{J}_{\phi}(\mathbf{{\mathbf{v}}})^\mathrm{T}\mathcal{J}_{\phi}(\mathbf{{\mathbf{v}}})\mathbf{e}d\mathbf{e}d{\mathbf{v}}\\
=&\int_{\mathcal{V}_\mathbf{v}}g(\mathbf{v})\mathbf{tr}\left[\frac{1}{V_{\mathbf{x}}}\int_{\mathcal{V}_\mathbf{x}}\mathbf{e}\mathbf{e}^\mathrm{T}d\mathbf{e}\mathcal{J}_{\phi}(\mathbf{{\mathbf{v}}})^\mathrm{T}\mathcal{J}_{\phi}(\mathbf{{\mathbf{v}}})\right]d{\mathbf{v}}\\
=&\int_{\mathcal{V}_\mathbf{v}}g(\mathbf{v})\mathbf{tr}\left[\frac{1}{K}D_\mathbf{x}\cdot \mathbf{I} \cdot \mathcal{J}_{\phi}(\mathbf{{\mathbf{v}}})^\mathrm{T}\mathcal{J}_{\phi}(\mathbf{{\mathbf{v}}})\right]d{\mathbf{v}}\\
=&\frac{1}{K}D_\mathbf{x}\cdot\int_{\mathcal{V}_\mathbf{v}}g(\mathbf{v})\mathbf{tr}\left[\mathcal{J}_{\phi}(\mathbf{{\mathbf{v}}})^\mathrm{T}\mathcal{J}_{\phi}(\mathbf{{\mathbf{v}}})\right]d{\mathbf{v}}\\
=&\frac{4}{K}D_\mathbf{x}\cdot\int_{\mathcal{V}_\mathbf{v}}g(\mathbf{v})\sum_{k=1}^Kv_kd{\mathbf{v}}\\
=&\frac{4}{K}D_\mathbf{x}\cdot\sum_{k=1}^K\int_{\mathcal{V}_{v_k}} \widetilde{g}(v_k)v_kd{v_k},
\end{split}
\end{equation}
where $\mathbf{I}$ is the identity matrix, the quantization noise $\mathbf{e}$ is white in the optimal lattices, $\widetilde{g}(v_k)$ is the marginal distribution of $v_k$, and we assumed that the quantization noise $\mathbf{e}$ is independent of $\mathbf{x}$ (and, therefore, independent of $\mathbf{v}$ as well)~\cite{Kleijn2010}. According to the neutrality~\cite{Ma2013} of the Dirichlet variable $\mathbf{v}$, the marginal distribution $\widetilde{g}(v_k)$ is beta distributed. Therefore, the mean value of $v_k$ with respect to its marginal distribution can be calculated explicitly. In our previous work, the measurement transformation between the LSF space and the $\Delta$LSF space was presented. Therefore, with these transformation methods, we can compare the high rate D-R performance in all the three different spaces fairly with consistent measurements.

\section{Statistical Model for SR$\Delta$LSF Parameters}
\label{sec: VMM}
The vMF distribution and its corresponding VMM are widely used in modeling the underlying distribution of the unit vector~\cite{Banerjee2005}. Therefore, we apply the VMM as the statistical model for SR$\Delta$LSF Parameters.
\subsection{Von Mises-Fisher Mixture Model}
Let $\mathbf{x}=[x_1,x_2,\ldots,x_k]^\mathrm{T}$ denote a $K$-dimensional vector satisfying $\sum_{k=1}^{K}x_k^2<1$. Then,
the $(K+1)$-dimensional unit random vector $[\mathbf{x}^{\text{T}}, 1-\sum_{k=1}^{K}x_k^2]^\mathrm{T}$ on the $K$-dimensional unit hypersphere $\mathbb{S}^{K}$ is said to have $(K+1)$-variate vMF distribution if its PDF is given by
 \begin{equation}\label{start}
 \eqs
\mathfrak{F}(\mathbf{x}\mid\boldsymbol{\mu},{\lambda})=\mathrm{c}_{K+1}(\lambda)~\mathrm{e}^{\lambda {\boldsymbol\mu}^{\mathrm{T}} \mathbf{x}},
 \end{equation}
where ${\parallel\boldsymbol\mu\parallel}=1$, $\lambda\geq0$, and $K\geq2$~\cite{Mardia2000}. The normalizing constant $\mathrm{c}_{K+1}(\lambda)$ is given by
\begin{equation} \label{nterm}
\eqs
\mathrm{c}_{K+1}(\lambda)=\frac{{\lambda}^
{\frac{K-1}{2}}}{{(2\pi)}^{\frac{K+1}{2}}~\mathcal{I}_{\frac{K-1}{2}}(\lambda)},
 \end{equation}
where $\mathcal{I}_{\nu} (\cdot)$ represents the modified Bessel function of the first kind of order $\nu$~\cite{Abramowitz1965}. The density function $\mathfrak{F}(\mathbf{x}\mid\boldsymbol{\mu},{\lambda})$ is characterized by the mean direction $\boldsymbol\mu$ and the concentration parameter $\lambda$.

With $I$ mixture components, the likelihood function of the VMM with~\emph{i.i.d.} observation $\mathbf{X}=[\mathbf{x}_1,\mathbf{x}_2,\ldots,\mathbf{x}_N]$ is
\begin{eqnarray}\label{vmm}
\eqs
f(\mathbf{{X}}\mid \mathbf{M}, \boldsymbol{\lambda},\boldsymbol{\pi} )=
\prod_{n=1}^{N}\sum_{i=1}^{I}\pi_i\mathfrak{{F}}(\mathbf{x}_n\mid \boldsymbol\mu_i, \lambda_i),
\end{eqnarray}
where $\boldsymbol{\pi}=[\pi_1,\pi_2,\ldots,\pi_I]^{\text{T}}$ ($\pi_i>0$, $\sum_{i=1}^{I}\pi_i=1$) is the weights, $\mathbf{M}=[\boldsymbol\mu_1,\boldsymbol\mu_2,\ldots,\boldsymbol\mu_I]$ is the mean directions, and $\boldsymbol{\lambda}=[\lambda_1,\lambda_2,\ldots,\lambda_I]^{\text{T}}$ is the concentration parameters.
\subsection{Parameter Estimation}

Let $\mathbf{{Z}}=\{\mathbf{z}_1,\mathbf{z}_2,\ldots,\mathbf{z}_I\}$ be the corresponding set of hidden random variables, where $\mathbf{z}_n=i$ means $\mathbf{x}_n$ is sampled from the $i$th vMF component. Given $\mathbf{{X}}$, $\mathbf{{Z}}$, and the model parameters $(\mathbf{M}, \boldsymbol{\lambda},\boldsymbol{\pi} )$, the complete log-likelihood of $\mathbf{{X}}$ writes
\begin{eqnarray}
\eqs
\ln p(\mathbf{{X}},\mathbf{{Z}}\mid \mathbf{M}, \boldsymbol{\lambda},\boldsymbol{\pi} )= \sum_{n=1}^{N}\ln \left[\pi_{\mathbf{z}_n}\mathfrak{{F}}(\mathbf{x}_n\mid \boldsymbol\mu_{\mathbf{z}_n}, \lambda_{\mathbf{z}_n})\right].
\end{eqnarray}
As obtaining the maximum-likelihood (ML) estimates from the complete log-likelihood is not tractable~\cite{Mardia2000}, an efficient expectation-maximization (EM) approach is developed which provides the ML estimates to the model parameters~\cite{Banerjee2005,Sra2012}. The E-step and the M-step are summarized as:
\begin{itemize}
  \item E-step \begin{eqnarray}
  \eqs
p(i\mid \mathbf{x}_n)=\frac{\alpha_i\mathfrak{{F}}(\mathbf{x}_n\mid {\boldsymbol\mu_i}, {\lambda_i})}{\sum_{j=1}^I\alpha_j \mathfrak{{F}}(\mathbf{x}_n\mid {\boldsymbol\mu_j}, {\lambda_j})}
\end{eqnarray}
  \item M-step
\eqs
  \begin{eqnarray}
\widehat{\alpha}_i&=&\frac{1}{n}\sum_{n=1}^{N}p(i\mid \mathbf{x}_n),\ \widehat{\boldsymbol\mu}_i=\frac{\sum_{n=1}^N\mathbf{x}_n p(i\mid \mathbf{x}_n)}{\|\sum_{n=1}^N\mathbf{x}_n p(i\mid \mathbf{x}_n)\|},\\
\bar r_i&=&\frac{\|\sum_{n=1}^N\mathbf{x}_n p(i\mid \mathbf{x}_n)\|}{\sum_{n=1}^Np(i\mid \mathbf{x}_n)},\ \widehat{\lambda}_i=\frac{\bar r_i K-{\bar r_i}^3}{1-{\bar r_i}^2}.
\end{eqnarray}
\end{itemize}

\section{PDF-optimized Vector Quantization}
\label{sec: PDF-optimized VQ}
In designing practical quantizers, one challenging problem is that when the amount of the training data is not sufficiently large enough, the obtained coodbook may tend to be over-fitted to the training set and perform worse for the whole real data set. The PDF-optimized VQ can overcome such problem either by generating sufficiently large amount of training data from the obtained PDF or calculating the optimal code book explicitly with the obtained PDF~\cite{Hedelin2000,Subramaniam2003}. Thus, with the trained VMM, we can design a PDF-optimized VQ.
\subsection{Distortion-Rate Relation with Constrained Entropy}
With the high rate assumption, the analysis of the quantization performance is analytically tractable~\cite{Kleijn2010}. Since coding at a finite rate is the motivation of using quanizers, constraint must be imposed on VQ design. Generally speaking, there are two commonly used cases, namely the constrained resolution (CR) and the constrained entropy (CE). In the CR case, the number of index levels is fixed. It is widely applied in communication systems. The CE case, on the other hand, imposes the constraint on average bit rate. It is less restrictive than the CR case and yields lower average bit rates. As the computational capabilities of hardware increases, it becomes more attractive to exploit advantages inherent in CE case~\cite{Kleijn2010}.

Assuming that the PDF of variable $\mathbf{x}$ is $f(\mathbf{x})$, the D-R relation in CE case, on a per dimension basis, writes
\begin{equation} \label{DR}
\eqs
D(R)=C(r,K)\cdot e^{-\frac{r}{K}\left(R-h(\mathbf{x})\right)},
 \end{equation}
 where $h(\mathbf{x})$ is the differential entropy of $\mathbf{x}$, $R$ is the average rate for quantization, and $C(r,K)$ is a constant depends on the distortion type $r$ (\emph{e.g.}, $r=2$ means the Euclidean distortion) and the variable's dimension (degrees of freedom) $K$.
\subsection{Optimal Inter-component Bit Allocation}
When applying a mixture model based quantizer, we model the PDF as a weighted addition of mixture components and design a quantizer for each component. The total rate $R$ will be divided into two parts, one for identifying the indices of the mixture components and the other for quantizing the mixture components. Given $I$ mixture components, the rate spent on identifying the indices is $R_{a}=\ln I$. The remaining rate $R_q=R-R_a$ will be used for quantizing the mixture components. Therefore, an optimal inter-component bit allocation strategy is required so that the designed quantizer can achieve the smallest mean distortion at a given $R_q$.

In CE case, the objective is to minimize the mean distortion
 \begin{equation}\label{ObjectiveFunc}
 \eqs
D(R) = \sum_{i=1}^I \pi_i D_i(R_i),
 \end{equation}
 where $R_i$ is the rate assigned to component $i$ and satisfies $R_q=\sum_{i=1}^I\pi_i R_i$. To reach the optimal mean distortion, each component should have its best CE performance. This indicates that the distortion for each mixture component writes
  \begin{equation}\label{CompDistor}
  \eqs
D_i(R_i)=C(r,K)\cdot e^{-\frac{r}{K}\left(R_i-h_i(\mathbf{x})\right)}.
 \end{equation}
The differential entropy for component $i$ in a VMM is
\begin{equation}\label{CompEntropy}
\eqs
\nonumber
\begin{split}
h_i(\mathbf{x})&= -\int \left[\ln\mathrm{c}_{K+1}(\lambda_i)+{\lambda_i {\boldsymbol\mu_i}^{\mathrm{T}} \mathbf{x}}\right]\cdot\mathrm{c}_{K+1}(\lambda_i)~\mathrm{e}^{\lambda_i {\boldsymbol\mu_i}^{\mathrm{T}} \mathbf{x}}d\mathbf{x}\\
&=-\ln \mathrm{c}_{K+1}(\lambda_i) - \lambda_i {\boldsymbol\mu_i}^{\mathrm{T}} \boldsymbol\mu_i\\
&=-\ln \mathrm{c}_{K+1}(\lambda_i) - \lambda_i,
\end{split}
\end{equation}
where we used the fact that $\|\boldsymbol\mu_i\|=1$.
\begin{figure*}[!t]

\psfrag{X}[][]{\tiny $R$ (in bit)}
\psfrag{Y}[][]{\tiny Distortion}
\vspace{-8mm}
     \centering
     \subfigure[\sps D-R performance of all VQs.\label{subfig:DRall}]{

          \includegraphics[width=.25\textwidth]{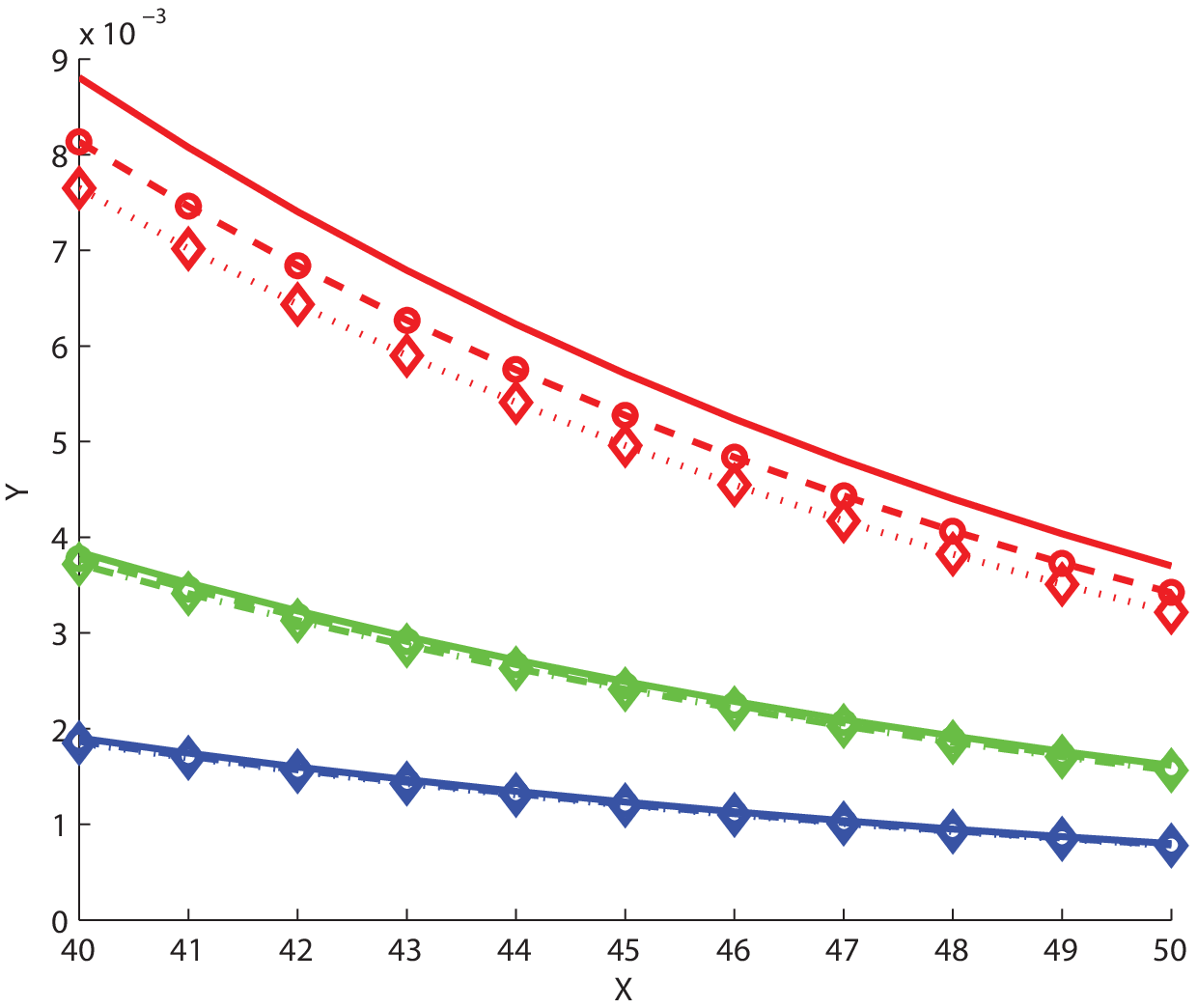}}
          \hspace{10mm}
     \subfigure[ \sps D-R relation for DVQ (zoomed in). \label{subfig:DRDVQ}]{

          \includegraphics[width=.25\textwidth]{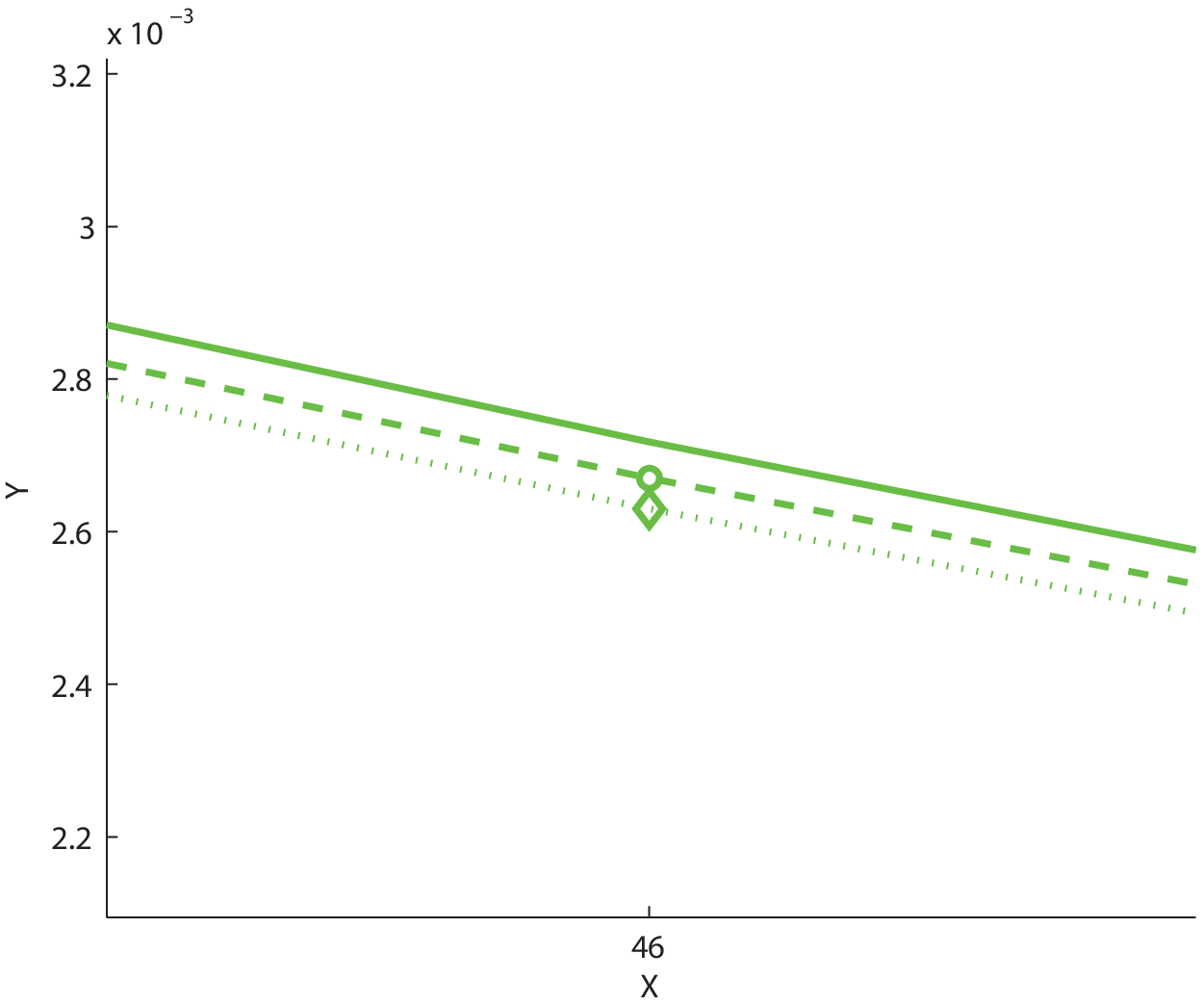}}
           \hspace{10mm}
     \subfigure[\sps D-R relation for VVQ (zoomed in).\label{subfig:DRVVQ}]{

          \includegraphics[width=.25\textwidth]{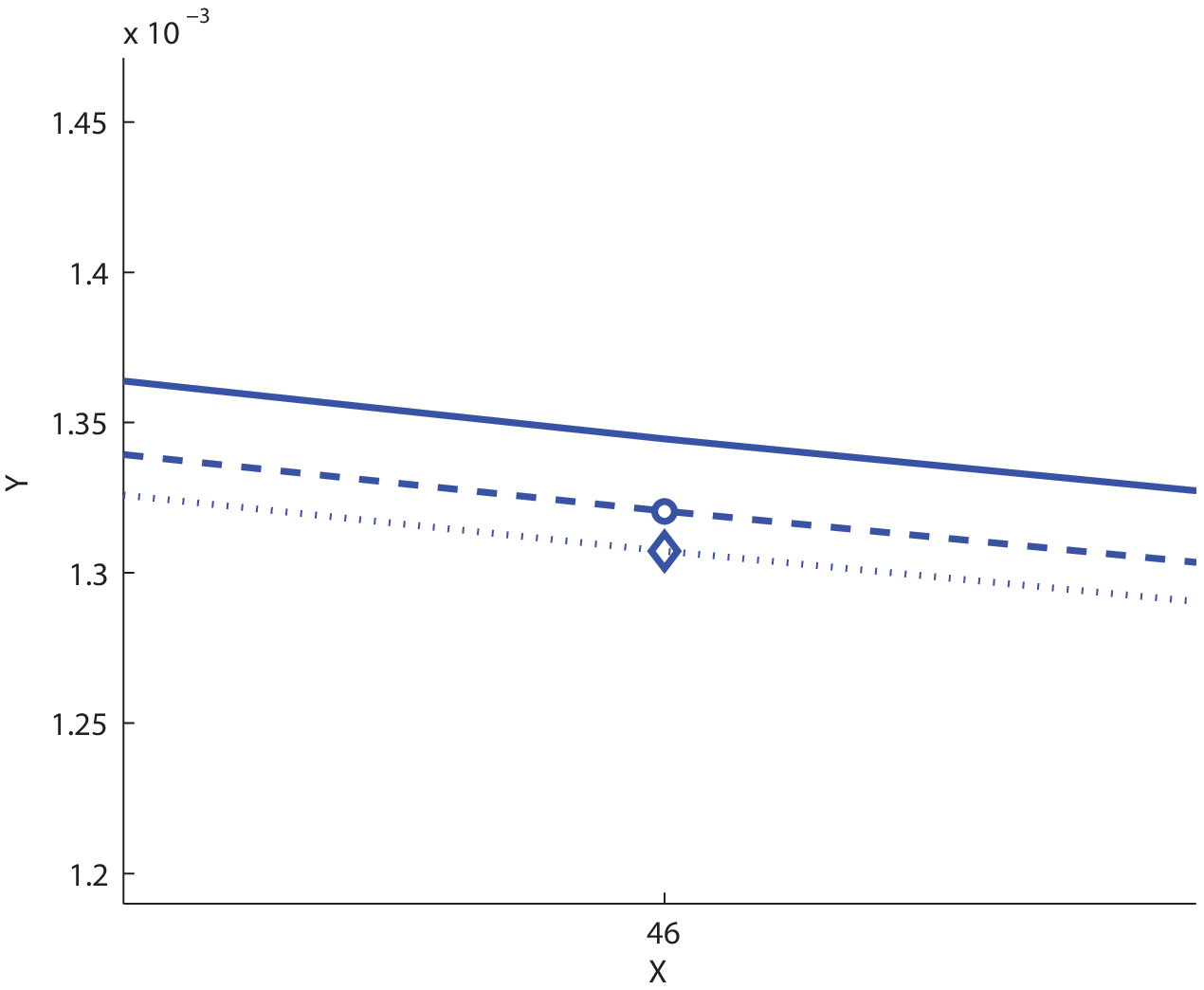}}
          \vspace{-2mm}
\caption{\sps D-R performance comparisons of GVQ, DVQ, and VVQ. To distinguish VQs, we use the red, green, and blue lines to denote the performance obtained by GVQ, DVQ, and VVQ, respectively. For each VQ, solid line, dash-circle line, and dot-diamond line represent the performance obtained by $16$, $32$, and $64$ mixture components, respectively.}\label{AAA}
          \vspace{-4mm}
\end{figure*}
The constrained optimization problem in~\eqref{ObjectiveFunc} can be solved by the method of Lagrange multipliers. With some mathematics, the rate assigned to the $i$th mixture component is
\begin{equation}\label{RateperComp}
\eqs
\begin{split}
R_i = R_q + h_i(\mathbf{x}) - \sum_{i=1}^I \pi_i h_i(\mathbf{x}).
\end{split}
\end{equation}
\subsection{Distortion-Rate Relation by VMM}
In CE case and with optimal inter-component bit allocation, the distortions contributed by all the mixture components are identical to each other because $R_i-h_i(\mathbf{x})$ is a constant which only depends on the trained model~\cite{Kleijn2010} . Then the D-R relation is
\begin{equation}\label{DRAll}
\eqs
\begin{split}
D(R)=\sum_{i=1}^I \pi_i D_i(R_i)=D_i(R_i),\ \ \forall i\in \{1,2,\ldots,I\}.
\end{split}
\end{equation}

\section{Experimental Results and Discussion}
\label{sec: Experiments}
The proposed inter-component bit allocation strategy optimizes the D-R relation of VVQ. To demonstrate the D-R performance, we compared it with our recently presented DVQ~\cite{Ma2013} and the widely used GVQ~\cite{Subramaniam2003,Chatterjee2008}. The TIMIT~\cite{TIMIT1990} database with wideband speech (sampled at $16$ kHz) was used. We extracted $16$-dimensional LPC parameters and transformed them to LSF parameters, $\Delta$LSF parameters, and SR$\Delta$LSF, respectively. With window length equal to $25$ milliseconds and step size equal to $20$ milliseconds, approximate $706,000$ LSF vectors (the same amount for $\Delta$LSF and SR$\Delta$LSF as well) were obtained from the training partition. GMM, DMM, and VMM were trained based on the relating vectors and the D-R relations were calculated, respectively. The mean values of $20$ rounds of simulations are reported. Figure~\ref{AAA} shows the D-R performance comparisons. It can be observed that VVQ leads to smaller distortion at different rates, compared to GVQ and DVQ. We believe this is due to the efficient modeling of the SR$\Delta$LSF parameters. Furthermore, better D-R performance can be obtained with more mixture components. Therefore, VVQ potentially permits superior practical VQ performance.
\section{Conclusion}
\label{sec: Conclusion}
A novel PDF-optimized VQ for LSF parameters quantization was proposed. The LSF parameters were transformed to the square-root $\Delta$LSF domain and we modeled the underlying distribution by a von Mises-Fisher mixture model (VMM). According to the principle of high rate quantization theory and with the constrained entropy case, the optimal inter-component bit allocation strategy was proposed based on the VMM. The mean distortion of the VMM based vector quantizer (VVQ) was minimized at a given rate so that the D-R relation was obtained. Compared to our recently proposed Dirichlet mixture model based VQ and the conventionally used Gaussian mixture model based VQ, the proposed VVQ performs better at a wide range of bit rates.
\section{Future Work}
\label{sec: Future Work}
For our future work, we need to implement a practical scheme to carry out the VQ. One possible solution is to propose an efficient quantizer for the von Mises-Fisher (vMF) source,~\emph{e.g.}, similar as the method in~\cite{Hamkins2002}. Another possible solution is to decorrelate the vMF vector variable into a set of scalar variables, each of which has an explicit PDF representation. Then we can replace the VQ with a set of independent scalar quantizers. This approach is similar to the Dirichlet source decorrelation and the Dirichlet mixture model based VQ introduced in~\cite{Ma2013}.
\vspace{0mm}

\normalsize
\appendices
\section{Discussion about the Inconsistency of Likelihood Comparison and D-R Comparison}
\emph{This section is only for discussion and will not appear in the final submission.}

As we observed before, the likelihood obtained by DMM is higher than the likelihood obtained by VMM. If we calculate the differential entropy of the trained PDF empirically as
\begin{equation}
\eqs
h(\mathbf{x})=-\mathbf{E}[\ln f(\mathbf{x})]\approx -\frac{1}{N}\sum_{n=1}^N \ln f(\mathbf{x}_n),
\end{equation}
a higher likelihood leads to a smaller differential entropy. According to~\eqref{DR}, this indicates better D-R performance. However, in our manuscript, VMM performs better than GMM, when we applied the mixture quantizer strategy.

\emph{Why would this happen?}

In our manuscript, for the CE case, we calculate the D-R performance of the mixture model by~\eqref{DRAll}. From~\eqref{DRAll}, we have
\begin{eqnarray}
\eqs
D(R)&=&D_i(R_i)\\
&=&C(r,K)\cdot e^{-\frac{r}{K}\left(R_i-h_i(\mathbf{x})\right)},\label{AA}\\
&=&C(r,K)\cdot e^{-\frac{r}{K}\left[\sum_{i=1}^I\pi_i\left(R_i-h_i(\mathbf{x})\right)\right]},\label{BB}\\
&=&C(r,K)\cdot e^{-\frac{r}{K}\left(R-\ln I - \sum_{i=1}^I\pi_i h_i(\mathbf{x})\right)}\label{CC}.
\end{eqnarray}
From~\eqref{AA} to~\eqref{BB}, we used the fact that $R_i-h_i(\mathbf{x})$ is the same for all $i$. From~\eqref{BB} to~\eqref{CC}, we used the fact that $R-\ln I=\sum_{i=1}^I R_i$.

Thereafter, we have
\begin{eqnarray}
\eqs
&&\ln I+\sum_{i=1}^I\pi_i h_i(\mathbf{x})\\
&\geq& -\sum_{i=1}^I\pi_i \ln\pi_i- \sum_{i=1}^I\pi_i \int \ln f_i(\mathbf{x})\cdot f_i(\mathbf{x})d \mathbf{x}\label{DD}\\
&=&-\sum_{i=1}^I\pi_i \int \ln\pi_i\cdot f_i(\mathbf{x})d\mathbf{x}- \sum_{i=1}^I\pi_i \int \ln f_i(\mathbf{x})\cdot f_i(\mathbf{x})d \mathbf{x}\nonumber\\
&=&-\sum_{i=1}^I\pi_i \int\ln\left[\pi_if_i(\mathbf{x})\right]\cdot f_i(\mathbf{x})d\mathbf{x}\\
&\geq&-\sum_{i=1}^I\pi_i \int\ln\left[\sum_{i=1}^I\pi_if_i(\mathbf{x})\right]\cdot f_i(\mathbf{x})d\mathbf{x}\label{EE}\\
&=&- \int\ln\left[\sum_{i=1}^I\pi_if_i(\mathbf{x})\right]\cdot \sum_{i=1}^I\pi_i f_i(\mathbf{x})d\mathbf{x}\\
&=&h(\mathbf{x}).
\end{eqnarray}

This inequality indicates that, the D-R performance calculated in the CE case (with mixture quantizer,~\eqref{DRAll}) is, in general, not identical to the D-R performance calculated with the whole PDF (\eqref{DR}). The inequality in~\eqref{DD} vanishes when all the components have the same weights. The equality~\eqref{EE} holds if there is no overlapping among the mixture components or we do not take the mixture modeling (I=1).

The inequality above introduces a systematic gap (a loss at the D-R performance). This gap depends on the training and the distribution assumption. Therefore, smaller differential entropy for the whole PDF ($h(\mathbf{x})$) can only guarantee better D-R performance, if we do not take mixture quantizer strategy. In mixture quantizer, it can not guarantee a better D-R performance.

\end{document}